\documentstyle[twocolumn,aps,epsfig,floats]{revtex}

\begin{document}

\draft
\tolerance = 10000

\setcounter{topnumber}{1}
\renewcommand{\topfraction}{0.9}
\renewcommand{\textfraction}{0.1}
\renewcommand{\floatpagefraction}{0.9}

\twocolumn[\hsize\textwidth\columnwidth\hsize\csname
@twocolumnfalse\endcsname

\title{Competing neural networks: \\ Finding a strategy for
the game of matching pennies}
\author{In\'es Samengo and Dami\'an H. Zanette}
\address{Consejo Nacional de
Investigaciones Cient\'{\i}ficas y T\'ecnicas\\ Centro At\'omico
Bariloche and Instituto Balseiro, 8400 Bariloche, Argentina}

\maketitle

\begin{abstract}
The ability  of a deterministic,   plastic system to learn  to imitate
stochastic  behavior is analyzed. Two  neural networks --actually, two
perceptrons-- are put  to play a  zero-sum game one against the other.
The competition, by acting as a kind  of mutually supervised learning,
drives  the networks  to produce  an   approximation  to the   optimal
strategy, that is to say, a random signal.
\end{abstract}

\pacs{PACS: 87.18.Sn, 02.50.Le, 05.45.-a, 05.45.Tp}

\vspace{1cm}

]

\section{Introduction}

Since the connection   between disordered spin systems and   symmetric
binary neural   networks was drawn \cite{Li74}  intensive theoretical,
numerical and experimental  research  has been  devoted to this  field
within physics,  and  in the  boundary of   physics with biology   and
information theory, among others \cite{Hz,NN}.   From the viewpoint of
the study of dynamical systems, neural  networks are a special kind of
distributed active systems \cite{Mikh}, which in their most impressive
realization --the brain--  are able to display extremely sophisticated
collective behavior.   Actual models have  of course much  more modest
scopes  but,  in spite of   their simplicity, they  have been  able to
imitate some basic features of cognitive processes.  These models have
also been  extended to perform  specific  tasks, such  as for instance
process control and forecasting \cite{fore}.

A basic capability of a wide class of neural-network models is that of
learning, i.e. the  possibility of modifying the internal architecture
of  the network to  adapt its dynamics   to an expected response. This
process  can take a variety  of forms, to  be  chosen according to the
aims of the model.   Pattern  storing and recognition  --the so-called
associative memory-- is perhaps  the best known \cite{assoc}.  Another
well known instance is  learning by generalization.  In this case, the
network   is exposed  to   some input information    and the output is
compared   with   the     expected   response.   Errors    are usually
backpropagated to modify the network  dynamics through a change in its
architecture. The network thus learns  from experience. It is expected
that after a certain learning transient the network is able to produce
the  correct output  even from  inputs  not included  in  the learning
sample. This kind of learning can be carried  on under supervision, or
the system can be designed to learn in an unsupervised manner, 
by means of a selforganization mechanism \cite{Hz,NN,Mikh}.

In this paper, we explore a neural-network  model of the learning that
takes   place during  a    competitive game.  Competitive   games have
recently attracted  a  great deal  of   attention among physicists  as
simple    models of  adaptive  evolution     and  selforganization  in
biological,  social,  and   economical  systems  \cite{game}.   Neural
networks have been designed and   trained to play some highly  complex
games such as chess and backgammon \cite{bg}.  The complexity of these
games, however,  does not allow a systematic  analysis of the learning
process  or a statistical evaluation  of the performance accomplished.
On the other hand, too simple games --such as those  that admit a pure
optimal strategy \cite{GT}-- should  be  readily solved by a  suitably
designed neural network.  In  fact,  finding  a pure strategy  can  be
associated with a maximization problem.

Here, we   focus the attention   at an intermediate level,  choosing a
competitive  zero-sum game with very simple  rules but  lacking a pure
optimal strategy,  v.g.   the game of   matching  pennies.  Two neural
networks are left to repeatedly play the game against each other.  The
successive  game   results are used  on-line   to  feed  the  learning
mechanism of the two players.   As in the case  of human players, each
network   tries to  guess  the strategy   of  its  opponent and, thus,
competition  becomes a   kind   of mutual supervision.    The  optimal
strategy for the game of matching pennies is  a purely stochastic one.
Thus,  the  challenge   for the networks,    whose  dynamics is  fully
deterministic, consists in approximating as close as possible a random
evolution.  Our analysis of the time series  generated during the game
shows that even small networks with simple architectures do quite well
--probably better than   any  human being  (not  using a   randomizing
device) \cite{MG}.

In  the  next section  we describe   in  detail the  game  of matching
pennies,  and specify the  architecture  and learning dynamics of  the
competing neural networks. Section III is devoted  to the study of the
model as a time-discrete dynamical system  --a mapping-- with emphasis
in its phase-space  evolution.  In Sect.  IV,  we analyze  statistical
properties of the dynamics during the game, evaluating the performance
of   the networks within  an  information-theory approach. Finally, we
discuss our results and consider some possible extensions.

\section{The game and the players}

In the  game of matching pennies,  player I chooses among two possible
instances, say ``heads'' or  ``tails.'' Player II, not  knowing player
I's choice, also chooses either ``heads'' or ``tails.''  Then, the two
choices   are disclosed --for example, each   player showing a penny--
and, if they are the same, player I  pays one cent to  player II.  If,
on the contrary, the choices have been different, II  pays one cent to
I.  The procedure is then repeated a large number of rounds, which has
for instance been defined by a previous agreement between the players.
In a less symmetric but very well-known  realization of the same game,
player II must guess in which  hand has player I hidden  a coin or any
other small object. The pay-off rules are the same as  for the game of
matching pennies. Since at each round player I's loss (or gain) equals
player II's gain (or loss), this is a zero-sum game. In game theory, a
two-player zero-sum game is said to be a ``strictly competitive'' game
\cite{GT}.

As the  game proceeds,  we expect  the two  players trying to outguess
each other,  keeping their  own strategies  secret.   Due to  the high
symmetry of the game of matching pennies, however, there is no optimal
pure strategy for either player.   Of course, it would be a  most poor
strategy for  any player  to choose  the same  instance at  every time
step. But, moreover, any deterministic way of deciding which  instance
should  be  chosen  at  a  given  time  step could be disclosed by the
opponent in  the long  run. On  the other  hand, trying  to guess  the
opponent's  strategy  could  lead  an  unsolvable, infinitely involved
problem.   As illustrated  in \cite{GT},  we may  picture player  I as
thinking: ``People usually  choose heads; hence  II will expect  me to
choose heads and choose heads  himself, and so I should  choose tails.
But perhaps II is reasoning along  the same line:  he'll expect  me to
choose tails, and so I'd better choose heads. But perhaps that is II's
reasoning, so...'' In this way,  it becomes impossible to determine  a
strategy in which either player could be confident. It follows that it
is necessary for both players to introduce a mixed stochastic strategy
where, at each time step,  each player chooses an instance  at random,
with a certain probability  distribution. The symmetry of  the present
game indicates clearly that the  best strategy for both players  is to
choose heads or tails with equal  probability.  In the long run,  this
insures a zero average gain, whereas any other strategy implies a  net
gain for the opponent.

Our aim here is to  study, as a dynamical  system, a pair of competing
neural networks playing the  game of matching pennies.  In particular,
we  are interested at analyzing whether  the dynamics implies learning
of an efficient   strategy ``on-line,'' i.e.    as the game  proceeds.
Since the  network dynamics and  the learning algorithm  considered in
the   following  are deterministic, it  cannot  be  expected  that the
networks will find  the  optimal (stochastic) strategy.   However,  it
could be possible that  the networks  were able  to approximate it  by
means  of  a complex deterministic dynamics   over a sufficiently long
period.  The basic  idea in the  learning process is  that the playing
strategy  of  each  network should  emerge  from trying  to  guess the
opponent's strategy.  This is in  fact the mechanism expected to drive
the game between human players: though a  general analysis of the game
shows that the best way of playing is at  random, each player tries to
outguess the other assuming a deterministic strategy, at least, in the
short  term.  The way  of playing derives   therefore from a (somewhat
paradoxical) cooperative   mechanism  during the contest,   where each
player ``supervises'' the learning of the other.

As for the architecture  of each neural network,  we take the simplest
model,  namely,   the perceptron,   introduced  in  \cite{Ros,Wid} and
reviewed in standard  books  on  neural  networks (see,   for example,
\cite{Hz,NN,I}).  It consists of a  collection  of $N$ inputs $s_i(t)$
and of $N$ synaptic weights $w_i(t)$ ($i=1,\dots ,N$) which define, at
each time step, a single output $\sigma(t)$ as
\begin{equation}    \label{perceptron}
\sigma(t) = S \left[\sum_i w_i(t) s_i (t)\right].
\end{equation}
Here  $S$  is  a  step-shaped  function,  that  we choose to be $S(x)=
\mbox{sign}(x)$. Thus, $\sigma=\pm 1$.  We associate each of  this two
possible values of the output with the instance chosen by the  network
at  a  given  time  step,  say,  $\sigma (t)=+1$ for heads and $\sigma
(t)=-1$ for tails.

\begin{figure}
\begin{center}
\psfig{figure=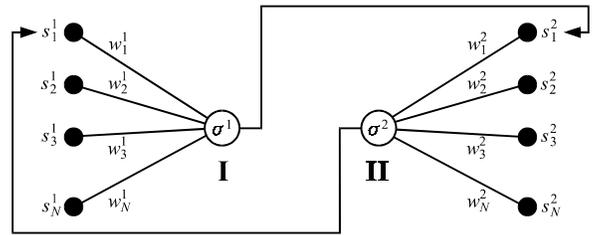,width=1.2\columnwidth}
\end{center}
\caption{Two competing perceptrons.}
\label{2percep}
\end{figure}

We  consider now two of  these perceptrons  (see Fig.  \ref{2percep}),
both with  $N$ inputs.  At each  time step, the  output of one  of the
perceptrons should be  determined by the  outputs of the  other at the
precedent  steps.  Indeed, this is the   information available to each
player on the  strategy of the  opponent.  We  associate therefore the
inputs $s_i^1$ of perceptron I with the previous outputs $\sigma^2$ of
perceptron II and {\it vice versa}, as
\begin{equation}    \label{inputs}
s_i^{1,2}(t)=\sigma^{2,1} (t-i) ,
\end{equation}
$i=1,\dots,N$. Time steps are of unitary length.

Learning is a consequence of the comparison  of the outputs of the two
perceptrons at each time.  If  the outputs are identical perceptron II
wins, and the synaptic weights $w_i^1$ of perceptron I are modified to
produce   a better  prediction of the   opponent's output  at the next
round.   Meanwhile, the synaptic weights  $w_i^2$ of perceptron II can
be left invariant, as they have led this perceptron  to win the round.
If,  on the other hand, the  outputs have  been different, $w_i^2$ are
modified  and $w_i^1$    are maintained.   A  suitable algorithm   for
implementing this mechanism  is the standard perceptron learning  rule
\cite{Hz,NN,I}, which in our case implies
\begin{equation}    \label{hebb1}
w_i^1(t+1)=w_i^1(t)-\eta \Theta [ \sigma^1(t) \sigma^2(t)]
 s_i^1(t) \sigma^2(t)
\end{equation}
and
\begin{equation}    \label{hebb2}
w_i^2(t+1)=w_i^2(t)+\eta \Theta [- \sigma^1(t) \sigma^2(t)]
\sigma^1(t) s_i^2(t) ,
\end{equation}
with $i=1,\dots , N$ and $\eta=(1+N)^{-1}$.
The Heaviside  function $\Theta$  --where $\Theta(x)=1$  for $x\ge  0$
and $\Theta(x)=0$ for  $x<0$-- acts here  as a mask,  by selecting the
perceptron whose synaptic weights are to be modified.

Suppose that the successive outputs of  perceptron I are replaced by a
periodic  series  of  $\pm  1$   \cite{JPA}.   From the   viewpoint of
perceptron II,  this is  interpreted  as the   opponent's choice of  a
trivial  strategy.   As a matter of  fact,  the perceptron convergence
theorem \cite{Hz,Ros,theor} insures that if $N$  is large enough, i.e.
if perceptron II's  memory  is sufficiently long-ranged, the  learning
procedure   stops and, from  then on,  perceptron  II wins all rounds.
When the period of  the output  series  of perceptron I is  lower than
$N$, in fact, it can be straightforwardly shown that there is at least
one  set of synaptic  weights $w^2_i$ that  make perceptron II able to
win  at  every round.  The  number  of steps  needed to  compute these
synaptic  weights  is of   order $N^3$  \cite{I},  and  can be  tested
numerically in our system.  It is therefore not expected that when two
large perceptrons are left  to play freely one of   them will adopt  a
short-period strategy.

\section{The system as a mapping: phase-space dynamics}

Equations (\ref{perceptron}) to  (\ref{hebb2}) define the  dynamics of
our  system.   They  can  be  resumed  in a $4N$-dimensional recursive
mapping for the perceptron inputs  and the synaptic weights only.  The
recursion equations are
\begin{equation}    \label{map}
\begin{array}{rl}
s_1^{1,2}(t+1)=&S[\sum_i w_i^{2,1}(t)s_i^{2,1}(t)], \\
s_i^{1,2}(t+1)=&s_{i-1}^{1,2}(t) \ \ \ (i=2,\dots,N), \\
w_i^1(t+1)=&w_i^1(t)-\eta \Theta [ s_1^1(t+1)s_1^2(t+1)]\\
&\times s_1^1(t+1) s_i^1(t) \ \ \ (i=1,\dots , N), \\
w_i^2(t+1)=&w_i^2(t)+\eta \Theta [-s_1^1(t+1)s_1^2(t+1)]\\
&\times s_1^2(t+1) s_i^2(t) \ \ \ (i=1,\dots , N).
\end{array}
\end{equation}
The phase space corresponding to  this mapping is discrete.   In fact,
the  inputs  $s_i^{1,2}$  can  adopt  the  two  values  $\pm  1$ only.
Moreover, $w_i^{1,2}$ can have real values but they vary on a discrete
set,  since  according  to  Eqs.  (\ref{hebb1})  and (\ref{hebb2}) the
variation of the synaptic weights has always the same modulus, $\left|
\Delta w^{1,2}\right| =\eta$.  Once the initial synaptic weights  have
been  fixed,  the  discrete  set  of  their  possible future values is
completely determined.

During the  evolution, the synaptic  weights can in principle run over
an infinite set. However, though the synaptic weights are not expected
to  converge to  fixed values but  to  continuously evolve as the game
proceeds,  it is reasonable to conjecture  that  they will not perform
arbitrarily long excursions in phase  space. To prove this conjecture,
let us consider in detail the evolution of the synaptic weights, given
by the two  last equations  in  (\ref{map}) or, equivalently, by  Eqs.
(\ref{hebb1}) and (\ref{hebb2}).  These two equations can be  written,
respectively, as
\begin{equation}    \label{w1}
\left\{
\begin{array}{ll}
w_i^1(t+1)=w_i^1(t)-\eta \sigma^1(t)s_i^1(t)
&\mbox{if $\sigma^1(t)=\sigma^2(t)$} \\
w_i^1(t+1)=w_i^1(t)
&\mbox{if $\sigma^1(t)=-\sigma^2(t)$}, \\
\end{array}
\right.
\end{equation}
and
\begin{equation}    \label{w2}
\left\{
\begin{array}{ll}
w_i^2(t+1)=w_i^2(t)-\eta \sigma^2(t)s_i^2(t)
&\mbox{if $\sigma^1(t)=-\sigma^2(t)$} \\
w_i^2(t+1)=w_i^2(t)
&\mbox{if $\sigma^1(t)=\sigma^2(t)$}. \\
\end{array}
\right.
\end{equation}
We now select one of the perceptrons and restrict the dynamics of  its
synaptic  weights  to  the  time  steps  where  they  are  effectively
modified, by  simply ignoring  the steps  where no  changes occur. The
evolution equations can be written in vectorial form as
\begin{equation}    \label{wvec}
{\bf w}(t+1)={\bf w}(t)-\eta S[{\bf w}(t)\cdot {\bf s}(t)] {\bf s}(t),
\end{equation}
where  the  components  of  ${\bf  w}$  and ${\bf s}$ are the synaptic
weights and the inputs of  the selected perceptron, respectively.   We
recall that  $S(x)$ is  the sign  function.   The scalar product ${\bf
w}\cdot{\bf   s}$   is   defined   in   the   usual   way,   cf.   Eq.
(\ref{perceptron}).    Note   that   (\ref{wvec})   holds   for   both
perceptrons.

Let us now consider for a moment that, in Eq. (\ref{wvec}), the vector
${\bf s}$ is independent of time. Under this assumption it is possible
to reduce the system (\ref{wvec}) to two equations for the  quantities
$p(t)={\bf s}\cdot {\bf w}(t)$ and $q(t)=|{\bf w}(t)|^2$, namely,
\begin{equation}    \label{pq}
\begin{array}{ll}
p(t+1)&=p(t)-(1-\eta) S[p(t)] \\
q(t+1)&=q(t)-2\eta | p(t)|+\eta(1-\eta).
\end{array}
\end{equation}
It can be  easily seen from  the first equation that $p(t)$ converges,
after a certain transient, to a period-2 cycle. The two values of  $p$
on this cycle, $p_1$ and $p_2$, satisfy the relation $p_2=p_1-1+\eta$.
They depend on  the initial conditions,  but are always  restricted to
the intervals $0<p_1<1-\eta$  and $\eta-1<p_2<0$. Accordingly,  $q(t)$
oscillates between two values, $q_1$ and $q_2$, defined by the initial
conditions and related by $q_2=q_1-2\eta p_1+\eta (1-\eta)$. After the
transient, the modulus of the vector ${\bf w}$ is therefore restricted
to vary within the interval $[-W,W]$ with $W =\max \left \{
\sqrt{q_1}, \sqrt{q_2} \right\}$.

In summary, for fixed $\bf s$ the evolution given by Eq.  (\ref{wvec})
drives  the  synaptic  weights  towards  a  bounded  domain whose size
depends  on  the  initial  condition  but  which is always finite.  We
stress that this is valid for  any choice of $\bf s$. Coming  now back
to the case of  variable inputs, we note  that the number of  possible
values for ${\bf  s}(t)$ is also  finite, and equals  $2^N$.  Equation
(\ref{wvec}) can therefore be thought  of as the application, at  each
time step, of  one of the  $2^N$ transformations just  studied.  Since
each of them contracts the space of synaptic weights towards a bounded
region, after the transient ${\bf w}(t)$ will always evolve within the
union of all those regions.  Disregarding transient effects, the space
of synaptic weights is then finite. Hence, the accessible phase  space
of mapping (\ref{map}) is finite and discrete.

\begin{figure}
\begin{center}
\psfig{figure=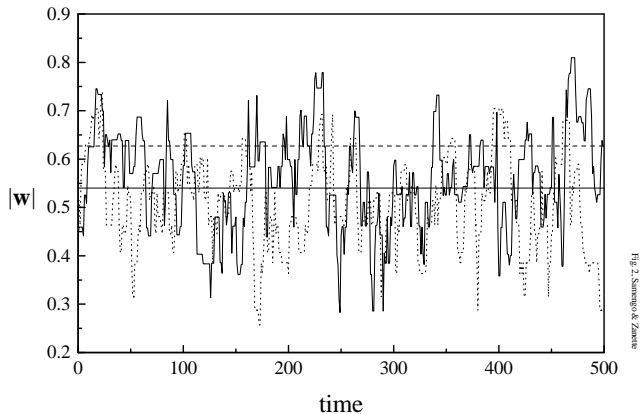,width=1.06\columnwidth}
\end{center}
\caption{
Module of the vector of synaptic weights for two competing perceptrons
with $N=10$ inputs,  as a  function of time.   Full  and dotted curves
correspond to $|{\bf w}^1|$ and $|{\bf  w}^2|$, respectively. Full and
dashed horizontal lines correspond to the analytical approximation for
the  average value   of  $|{\bf  w}|$  for $N=10$   and $N\to \infty$,
respectively.
}
\label{modulw}
\end{figure}

As an illustration  of the evolution  of synaptic weights,  we show in
Fig.   \ref{modulw}  the  time  dependence  of  $|{\bf  w}|$  for both
perceptrons. The initial  weights were uniformly  chosen at random  in
$(-0.2,0.2)$, and $N=10$. The horizontal  lines in the plot stand  for
the theoretical  values of  the temporal  average of  $|{\bf w}|$  for
$N=10$  (full  line)  and  $N\to  \infty$  (dashed line). These can be
calculated by taking the square of Eq. (\ref{wvec}), namely,
\begin{equation}    \label{q}
q(t+1)=q(t)-2\eta \sqrt{q(t)} | \widehat {\bf w}(t) \cdot {\bf
s}(t)|+\eta(1-\eta)
\end{equation}
with $\widehat {\bf w}={\bf w}/|{\bf w}|$. This recursion  equation is
analogous to the second of Eqs.  (\ref{pq}). It can be seen that,  for
sufficiently large $N$, the average of $|\widehat{\bf w}(t) \cdot {\bf
s}(t)|$ over time --or, equivalently, over random realizations of  the
vectors $\widehat {\bf  w}$ and $\bf  s$-- becomes independent  of $N$
and  approaches  the  limit  $\langle  |  \widehat{\bf  w}\cdot   {\bf
s}|\rangle  =  \sqrt{2/\pi}\approx  0.798$.  From  Eq. (\ref{q}), this
implies that for large $N$:
\begin{equation}    \label{avew}
\langle |{\bf w}|\rangle = \langle \sqrt{q} \rangle=
\sqrt{\frac{\pi}{8}} \approx  0.627,
\end{equation}
cf. \cite{NC}.
A better approximation  for finite $N$  is $\langle |{\bf  w}| \rangle
=\sqrt{\pi N(N-1)/8(N+1)^2}$. For $N=10$ this gives $\langle |{\bf  w}
|\rangle  \approx  0.540$,  which  is   the  value  plotted  in   Fig.
\ref{modulw}. The average  value of $|{\bf  w}|$ provides an  estimate
for the size of the domain  of phase space where the synaptic  weights
evolve after transients have elapsed. Note that the fact that $\langle
|{\bf w} |\rangle$ approaches a  constant for large $N$ implies  that,
in average, the synaptic weights are $w_i^{1,2} \sim 1/\sqrt{N}$.

\begin{figure}
\begin{center}
\psfig{figure=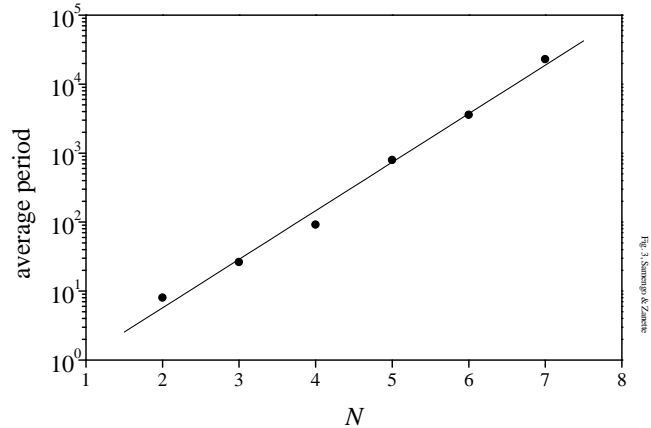,width=1.08\columnwidth}
\end{center}
\caption{
Average periods of the orbits   of mapping (\ref{map}), for  different
values of $N$. Notice the logarithmic scale in the vertical axis.
}
\label{periods}
\end{figure}

The main  byproduct of  the fact  that for  our system  phase space is
finite and  discrete is  that, after  the transient  has elapsed,  the
orbits will be  periodic. It becomes  therefore relevant to  determine
the length of  the periods. In  fact, if it  resulted that orbits  get
typically  trapped  in  short  cycles,  the  problem would at once get
uninteresting. We have measured the periods numerically, carrying  out
extensive  series  of  $100$  to  $1000$ realizations $6.4\times 10^5$
steps long, with $N$ ranging from $2$ to $10$. Initial conditions were
chosen at random, with  the synaptic weights uniformly  distributed in
$(-0.2,0.2)$. The  system has  always been  found to  reach a periodic
orbit for $N<7$.  For a  fixed value of $N$, periods show  typically a
broad  distribution.  The  average  period  has been found to increase
exponentially with $N$,  as shown in  Fig.  \ref{periods}.   For $N\ge
7$, not  all the  realizations displayed  periodicity, indicating  the
occurrence of  periods longer  than our  numerical realizations.  Some
test realizations for  $N=10$ suggest that  periods could grow  beyond
$10^8$ steps.

The system thus seems to have two well-differentiated time scales.  On
the one hand,  there should be a time  scale associated with learning,
of  order $N^3$. As stated  above, in the  case of a single perceptron
being trained to predict a periodic series this  is in fact the number
of steps  needed  to compute  all  the   synaptic weights.   For   the
competing  perceptrons,  the length  of the   initial transient during
which the system explores phase space to find the bounded region where
it will evolve later, should be of the same  order. On the other hand,
we have   a much  longer   ``recursion'' time  scale,  of  order $A^N$
($A\approx 5.05$, Fig. \ref{periods}), associated  with the periods of
orbits inside that   region.  Though the two-perceptron  dynamics   is
dissipative, it resembles in this  aspect that of Hamiltonian  systems
with  many degrees of  freedom.    Indeed, according to   Poincar\'e's
theorem \cite{Poinc},  Hamiltonian   systems  are recurrent and,    at
sufficiently long times, they visit  an arbitrarily small neighborhood
of their  initial  state.  However, in   a statistical description  of
their evolution, it is possible to  identify much shorter time scales,
related to the relaxation of fast variables \cite{fast}.

At   the   level  of recursion   time  scales,  the   dynamics  of the
two-perceptron  system is  in a  sense  trivial.  Orbits  are  in fact
periodic at long times, and the results of successive game rounds will
be repeated {\it  ad infinitum}.  When, during  a whole period, one of
the perceptrons is  able to gain  even the smallest advantage over the
other,  this small difference  will continuously accumulate producing,
in the long run, an arbitrarily large bias in  the result of the game.
As in the case of large Hamiltonian  systems, however, recursion times
are far beyond the reach of our (numerical) experience  as the size of
the perceptrons increases. Therefore, most of  the realizations of the
two-perceptron game analyzed bellow will  always be restricted to  the
transient period, previous to the appearance  of periodicity.  In this
stage, the relevant time  scale is the learning  time, of order $N^3$.
Within such times we  expect the system to  reach a kind of stationary
playing  regime where,  if the  learning  algorithm is efficient,  the
outputs of the two perceptrons should imitate a  random series of $\pm
1$. In the next section we study the statistical properties of these
output series.

\section{Statistical analysis of the game dynamics}

Random properties in time series can be  characterized in a variety of
ways. In our case, where the relevant series are  arrays of $\pm 1$, a
suitable measure of time  correlations is an information-like quantity
\cite{Treves}. As shown    below, this   quantity   can be   used   to
characterize   the    correlation   between   different   series  and,
consequently, the correlation  of a series   with itself.  It  has the
advantage of  being    additive, and is  therefore   appropriate  when
comparing  numerical results. We  thus begin  by  defining  the mutual
information of two time series.

Consider two dichotomic stochastic processes  $S_1$ and $S_2$ that, at
each time step, can adopt the values  $\pm 1$ with certain probability
distributions.    Let $P(S_1,S_2)$ be the   joint  probability for the
processes,  and  $P_1(S_1) =  \sum_{S_2}   P(S_1,S_2)$ and $P_2(S_2) =
\sum_{S_1}   P(S_1,S_2)$    their  individual    (marginal  \cite{VK})
probabilities.  A measure of the correlation between the two processes
is given by the {\it mutual information} \cite{Treves}, defined as
\begin{equation}    \label{mut}
I= \sum_{S_1=\pm 1}\sum_{S_2=\pm 1} P(S_1,S_2) \log_2 \left[
P(S_1,S_2)\over P_1(S_1) P_2(S_2)\right].
\end{equation}
It can be shown that $I \ge 0$. For two uncorrelated processes,  where
$P(S_1,S_2)=P_1(S_1)P_2(S_2)$,  the  mutual  information  reaches  its
minimum,  $I=0$.  The  maximal  value  of  the  mutual  information is
obtained for two identical stochastic processes, $S_1=S_2$, where  $I=
-P_1(+1)  \log_2P_1(+1)-P_1(-1)  \log_2P_1(-1)$.  In  particular,   if
$P_1(+1)=P_1(-1)=1/2$, we get $I=1$.

The definition of   mutual  information, Eq.  (\ref{mut}),    suggests
immediately a way  of introducing a  measure of autocorrelation  for a
single dichotomic stochastic process $S$ at different times.  In fact,
associating  $S_1(t)$  and  $S_2(t)$  with   $S(t)$ and   $S(t+\tau)$,
respectively, we can introduce the (two-time) {\it autoinformation} as
\begin{equation}    \label{auto}
\begin{array}{rl}
I(t,\tau)=\sum_{S(t)}\sum_{S(t+\tau)}& P[S(t),S(t+\tau)] \\
\\ &\times
\log_2 \left\{ P[S(t),S(t+\tau)]\over P[S(t)] P[S(t+\tau)]\right\} .
\end{array}
\end{equation}
If   $S$   is   a   stationary   stochastic   process   \cite{VK}  the
autoinformation depends  on the  time interval  $\tau$ only,  $I\equiv
I(\tau)$.  If  the successive values  of $S$ are  uncorrelated we have
$I=0$, whereas for $\tau=0$ we  get the maximal value $I(t,0)=  -P(+1)
\log_2P(+1)-P(-1) \log_2P(-1)$.

In practice, for a finite realization of the stochastic processes, the
probabilities  involved  in  Eqs.   (\ref{mut})  and  (\ref{auto}) are
approximated by the corresponding  frequencies, which can be  computed
by simple  counting of  the relevant  occurrences. This  approximation
implies that  in the  case of uncorrelated processes   the information
can differ from zero, due  to fluctuations in the finite  sample under
consideration. It  can be  shown that  for a  $T$-step realization  of
uncorrelated stochastic processes  where the individual  probabilities
of the two possible values  $\pm 1$ are equal, $P(+1)=P(-1)=1/2$,  the
probability distribution for the information to have a value $I$ is
\begin{equation}    \label{I}
p_T(I)= \sqrt{ T\ln 2 \over \pi I } \exp(-TI\ln 2),
\end{equation}
for small $I$. The resulting mean value of the information is
\begin{equation}    \label{meanI}
\langle I\rangle =\int_0^\infty Ip_T(I) dI= {1\over 2T\ln 2},
\end{equation}
which decreases as $T^{-1}$ as  the series size grows. For  large $T$,
$p_T(I)\to \delta (I)$, as expected. Thus, the distribution of  values
for the information computed from  finite samples of size $T$  and its
average are  to be  respectively compared  with $p_T(I)$  and $\langle
I\rangle$ in order to detect the presence of correlations.

We now  consider two  playing perceptrons  with $N=10$,  and apply the
definition of autoinformation (\ref{auto}) to any of the two series of
outputs, $S(t)\equiv \sigma^{1,2}(t)$.  The outputs are recorded after
the first $10^4$ steps have  elapsed, in order to avoid  nonstationary
transient effects during the first  stage of learning (of order  $N^3$
\cite{I}).  The  recorded  series  are  $T=10^4$  steps  long, and the
results presented  below correspond  to averages  over $5\times  10^4$
realizations.

\begin{figure}
\begin{center}
\psfig{figure=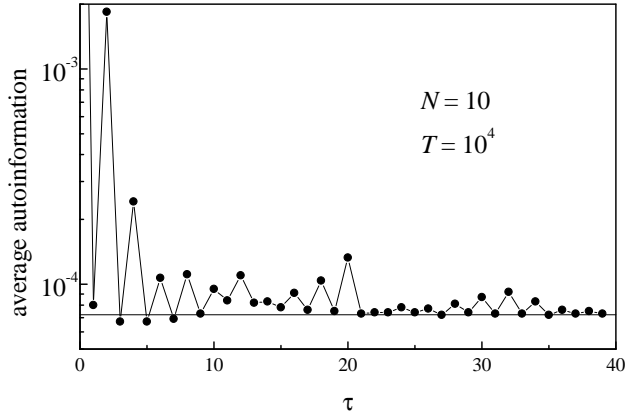,width=1.08\columnwidth}
\end{center}
\caption{
Average  autoinformation for  the  output  of  one of   the
competing perceptrons   as a function   of the time   interval $\tau$,
measured in series of $10^4$ steps. Both perceptrons have $10$ inputs.
The horizontal  line   corresponds to   the average    autoinformation
expected for an uncorrelated series of the same length.
}
\label{avinfor}
\end{figure}

Figure \ref{avinfor} shows the  measured average autoinformation as  a
function of  $\tau$. The  horizontal line  corresponds to  the average
autoinformation (\ref{meanI}) expected for an uncorrelated series with
the present value  of $T$, i.e.  $\langle I\rangle \approx  7.2 \times
10^{-5}$.   We  first  note  that,  except  for  $\tau=2$ and $4$, the
autoinformation of  the output  signal is  always less  that twice the
value of $\langle  I\rangle$ for a  random series.   This implies that
each  perceptron  exhibits  a  quite  good performance in generating a
random  sequence.   There  are  however  certain regular patterns that
suggest the presence of small but nontrivial correlations. Indeed, the
average autoinformation oscillates strongly for small $\tau$, reaching
high levels for  even values of  $\tau$ and dropping  abruptly for odd
values of $\tau$.  On  average, these oscillations decrease as  $\tau$
grows, but they  reappear near $\tau=20$  and $30$.   Realizations for
other values of $N$ indicate that the oscillation amplitude  decreases
as $N$ grows, and that  the ``bursts'' at which oscillations  reappear
occur when $\tau$ approaches integer multiples of $N$.  The  amplitude
of these bursts decreases for larger multiples.

\begin{figure}
\begin{center}
\psfig{figure=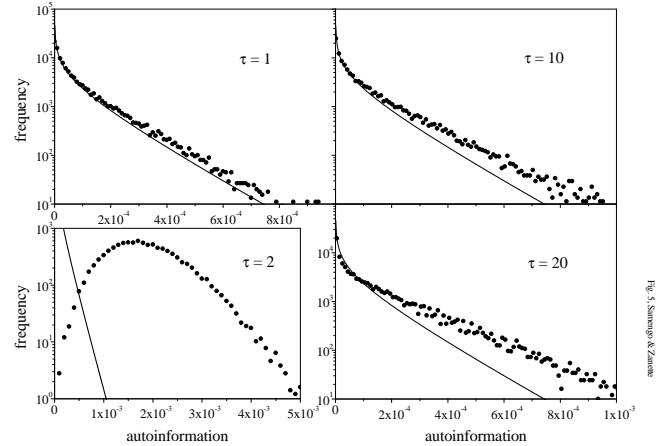,width=1.08\columnwidth}
\end{center}
\caption{
Normalized  frequencies   of autoinformation values   obtained from $5
\times 10^4$ series  of $10^4$ steps,  for several values of the  time
interval $\tau$. Both perceptrons have $10$ inputs.  Curves correspond
to the   distribution expected for  uncorrelated  series  of  the same
length, Eq. (\ref{I}).
}
\label{hist}
\end{figure}

A more detailed  description of the appearance  of correlations in the
output signals of the perceptrons  is provided by the distribution  of
autoinformation  values.   Figure \ref{hist}  displays the  normalized
frequencies of autoinformation values   resulting from our sets  of $5
\times 10^4$ realizations of $10^4$-step series  for various values of
$\tau$.  The curve corresponds to $p_T(I)$ for an uncorrelated series,
Eq. (\ref{I}). For  $\tau=1$ practically no  correlations are detected
by the autoinformation. We note only a slight overpopulation for large
$I$. On the other hand, for $\tau=2$, which corresponds to the largest
deviation in the average autoinformation (see Fig. \ref{avinfor}), the
distribution is qualitatively different.  It  exhibits a maximum at  a
rather large value of the autoinformation ($I\approx 2\times 10^{-3}$)
and, except for small values of $I$,  it is systematically much larger
than the distribution expected for a random series.  At $\tau=10$, the
distribution has a profile similar  to that observed for $\tau=1$, but
the  overpopulation  at  the     tail is noticeably   larger.     This
overpopulation  grows  further during  the bursts   where oscillations
reappear.  The plot for $\tau=20$ shows  the distribution at the first
of these bursts. In contrast, for the intermediate values at which the
average autoinformation plotted   in  Fig. \ref{avinfor}  reaches  the
information of a  random series, the corresponding distribution cannot
be distinguished from $p_T(I)$.

We have found   that the oscillations  of the  average autoinformation
shown in  Fig.  \ref{avinfor}   are  essentially a byproduct    of the
internal dynamics of each perceptron. In fact, if instead of using the
opponent's output, a perceptron  is fed with a  random series of  $\pm
1$,  the autoinformation  of its  own  output  oscillates as  well.  A
detailed analysis of the output  series reveals that, for even $\tau$,
the product $\sigma(t)   \sigma(t+\tau)$ is  more frequently  negative
than positive. For instance,  for $\tau=2$, the respective frequencies
are  about $0.52$ and $0.48$.   We remark in  passing  that this small
relative  difference --of  the order  of a  few percent--  produces an
increment larger than  one order of  magnitude in the autoinformation,
which evidences the  sensibility of   this  quantity as a measure   of
correlations.  For larger  values  of $\tau$, the  difference  is even
smaller.    On the  other  hand, for   odd  $\tau$ no  differences are
detected.

In order to  trace the origin  of  the correlations observed  for even
$\tau$, a careful analysis of the learning algorithm has to be carried
out. We consider first the case of $\tau=2$. After two time steps, the
vector of synaptic weights can be written as
\begin{equation}    \label{wt2}
\begin{array}{rl}
{\bf w}(t+2)=&{\bf w}(t)-\eta \theta(t) \sigma(t) {\bf s}(t) \\
&-\eta \theta(t+1) \sigma(t+1) {\bf s}(t+1),
\end{array}
\end{equation}
where $\theta(t)=1$ if the weights have been modified at time $t$, and
$\theta(t)=0$ otherwise  [cf. Eq. (\ref{wvec})].   When the perceptron
if fed  with a random signal, $\theta(t)$  can be seen as a stochastic
process with  equal probabilities for  its two  values. The product of
the outputs two steps apart is
\begin{equation}    \label{d1}
\begin{array}{lll}
\sigma(t) \sigma(t+2)&=&\sigma(t) S[{\bf w}(t+2)\cdot {\bf s}(t+2)] \\
&=&S[\sigma(t) {\bf w}(t)\cdot {\bf s}(t+2)-\eta \theta(t) {\bf 
s}(t)\cdot {\bf s}(t+2)\\
& &-\eta \theta(t+1)\sigma(t) \sigma(t+1) {\bf 
s}(t+1)\cdot {\bf s}(t+2)]
\end{array}
\end{equation}
Numerical   measurements  of the  right-hand   side  (r.h.s.) of  this
equation show  that the  first two terms  in the  argument of the sign
function have zero mean and  do not produce a  net contribution to the
sign   of  $\sigma(t)\sigma(t+2)$.    The  only  contribution   to the
correlation is  originated in the  third term.   To  verify this  fact
analytically, we first note that
\begin{equation}    \label{d2}
\begin{array}{l}
{\bf w}(t+1)={\bf w}(t) [1+{\cal O}(1/\sqrt{N})], \\
{\bf s}(t+1)\cdot {\bf s}(t+2)=
{\bf s}(t)\cdot {\bf s}(t+1) [1+{\cal O}(1/\sqrt{N})].
\end{array}
\end{equation}
The first of these identities results from the fact that, as shown  in
the previous section, $w_i \sim 1/\sqrt{N}$ whereas, according to  Eq.
(\ref{map}), its  variation in  one time  step is  given by  $\eta\sim
1/N$. The second identity can be readily proven from the evolution  of
$s_i(t)$, also  given in  Eq.   (\ref{map}). Consequently,  neglecting
terms  of  order  $1/\sqrt{N}$,  the  sign  of  the product $\sigma(t)
\sigma(t+1) {\bf  s}(t+1)\cdot {\bf  s}(t+2)$ can  be approximated  as
follows:
\begin{equation}    \label{d3}
\begin{array}{l}
S[\sigma(t) \sigma(t+1) {\bf s}(t+1)\cdot {\bf s}(t+2)]\\ \approx
S\{ [ {\bf w}(t)\cdot {\bf s}(t)][ {\bf w}(t)\cdot {\bf s}(t+1)][
{\bf s}(t)\cdot {\bf s}(t+1)]\} .
\end{array}
\end{equation}
Note that the  argument of the  sign function in  the r.h.s.   of this
equation is likely to be positive, since it is given by the product of
the projections of two vectors, ${\bf s}(t)$ and ${\bf s}(t+1)$, along
the direction of ${\bf w}(t)$ times their mutual scalar product.  More
explicitly,
\begin{equation}    \label{d4}
\begin{array}{l}
[ {\bf w}(t)\cdot {\bf s}(t)][ {\bf w}(t)\cdot {\bf s}(t+1)][
{\bf s}(t)\cdot {\bf s}(t+1)]\\ = s_w^2(t)s_w^2(t+1)+
s_w(t)s_w(t+1) {\bf s}'(t)\cdot {\bf s}'(t+1),
\end{array}
\end{equation}
with  $s_w={\bf  w}\cdot   {\bf  s}$  and   ${\bf  s}'={\bf  s}-   s_w
\widehat{\bf w}$.  The first term  in the r.h.s.  of this  equation is
always positive,  whereas the  second term  is not  expected to have a
definite sign on average. Note moreover that the first term is of  the
order of unity, whereas the  second term is of order  $\sqrt{N}$. This
implies  that  the  relative  importance  of the positive contribution
decreases as $N$ grows.   Coming now back to  Eq.  (\ref{d1})  through
Eq(\ref{d3}) it is clear that, when the synaptic weights are  modified
at time $t+1$ (i.e. $\theta(t+1)=1$),  there is in average a  negative
contribution to  $\sigma(t)\sigma(t+2)$, in  agreement with  numerical
results.  According  to  the  above  analysis, this correlation should
become less important as $N$ grows.  In fact, the autoinformation peak
at $\tau=2$ is observed to decrease in the simulations.

For  arbitrary  $\tau$,  the  analysis  can  be  repeated {\it mutatis
mutandis}. We have
\begin{equation}    \label{d5}
\begin{array}{rl}
\sigma(t) \sigma(t+\tau)=S[& \sigma(t) {\bf w}(t)\cdot {\bf 
s}(t+\tau)\\
&-\eta \sum_{t'=0}^{\tau-1} \theta(t+t') \sigma(t) 
\sigma(t+t')\\ &\times {\bf s}(t+t')\cdot {\bf s}(t+\tau)] .
\end{array}
\end{equation}
Taking now into account that
\begin{equation}    \label{d6}
\begin{array}{l}
{\bf w}(t+t')={\bf w}(t) [1+{\cal O}(\sqrt{t'/N})], \\
{\bf s}(t+t')\cdot {\bf s}(t+\tau)=
{\bf s}(t)\cdot {\bf s}(t+\tau-t') [1+{\cal O}(\sqrt{t'/N})],
\end{array}
\end{equation}
the sign of  the product $\sigma(t)  \sigma(t+t') {\bf s}(t+t')  \cdot
{\bf s}(t+\tau)$  in the  sum of  Eq. (\ref{d5})  can be approximately
written as
\begin{equation}    \label{d7}
\begin{array}{l}
S[\sigma(t) \sigma(t+t') {\bf s}(t+t')\cdot {\bf s}(t+\tau)]\\
\approx
S\{ [ {\bf w}(t)\cdot {\bf s}(t)][ {\bf w}(t)\cdot {\bf s}(t+t')][
{\bf s}(t)\cdot {\bf s}(t+\tau-t')]\} .
\end{array}
\end{equation}
The argument of the sign function in the r.h.s. of this equation has a
positive contribution of the same type as in Eq. (\ref{d3}) when $t+t'
=t+\tau-t'$, i.e. for $\tau=2t'$. Therefore, in the realizations where
$\theta(t+\tau/2)=1$   a   negative    contribution   to    $\sigma(t)
\sigma(t+\tau)$ appears. This  of course requires  $\tau$ to be  even.
Since other contributions have no definite sign, peaks in the  average
autoinformation are expected for  even values of $\tau$,  as observed.
Note moreover that the order  $\sqrt{t'/N}$ of the terms neglected  in
Eq. (\ref{d7}) increases  with $t'$, i.e.  with $\tau$. This  explains
why the height of the peaks decreases as $\tau$ grows. 

Along the same line of analysis, it is  possible to explain the bursts
where the    autoinformation  peaks reappear.   Now, however,   it  is
necessary to take into  account both perceptrons.  In fact, the output
of a single perceptron fed with a random  signal does not exhibit such
bursts. They  are rather a  consequence of the interaction between the
two perceptrons  during the game.  The analysis, whose details we omit
here, shows that bursts are originated by a kind of bouncing effect in
the transmission of information  between the opponents.  This bouncing
effects  is  attenuated as $\tau$    grows, and decreases  for  larger
perceptrons, as observed in the numerical simulations.

In summary, the  statistical analysis of   perceptron outputs at  time
scales  larger   than the  learning stage   but much  shorter than the
recursion times, reveals    that the perceptrons  are quite  efficient
players of the game of matching pennies. Even  with a relatively small
number of inputs,  i.e. with a  relatively short-ranged memory,  their
dynamics is able to generate quasi-random mixed strategies.  We recall
that this  behavior originates   spontaneously from the  deterministic
learning  algorithm with which each  player is endowed to outguess its
opponent.   Remaining  correlations,   which  could   in principle  be
exploited by a ``smarter''  opponent to obtain  a net gain during  the
game, are overall small  and can in fact  be reduced systematically by
increasing the memory range.

\section{Discussion}

We have  here considered an  example of a fully deterministic learning
system and explored its ability to behave stochastically.  Concretely,
we have coupled two deterministic perceptrons in such  a way that they
imitate  two players  of   the game  of matching  pennies,   trying to
outguess each other.  Since the optimal strategy  for  this game is  a
purely stochastic   sequence of outputs,   the learning process should
lead the network dynamics to approach a random signal.

In the first  place, we have  observed  that a perceptron producing  a
periodic signal can always   be defeated by a  sufficiently  ``smart''
opponent, i.e.  by a  perceptron with a  sufficiently large  number of
neurons. This kind of  ``dummy''  player provides  in fact a  linearly
separable set of examples for the learning  of its opponent \cite{Hz}.
The  learning task is thus   to find a  plane  in the input space that
separates  the input  states  into  two  groups,  namely those   whose
expected outputs are either $+1$ or $-1$. On the  other hand, when the
two competing perceptrons are allowed to learn the situation is pretty
much different.  Since  both networks   are   looking  for  the   best
performance, they both change their strategies on line and, thus, they
may well provide not only a nonlinearly separable set of examples, but
also an  inconsistent one. That is to  say, at two different times any
perceptron can give two  different outputs from  the same input state.
This  is   the reason why  the  learning  process  does, in  fact, not
converge,   and why the system    is expected to spontaneously develop
stochastic-like dynamics.

Our main conclusion  is that,   despite  the  fact that the    overall
dynamics is in the   long run periodic,  the  perceptrons do learn  to
behave quasi-stochastically over moderately  long time intervals.   An
information-theoretical   statistical  analysis of  the output signals
shows slight time correlations,   to be ascribed to  the deterministic
coupling between the learning  mechanism  and the outputs  themselves,
which act as  the inputs of  the respective opponents.   The effect of
these correlations is observed to  decrease gradually as the number of
neurons in each perceptron grows. Two seemingly paradoxical aspects of
this  learning  process deserve to  be pointed  out,  because of their
suggestive similarity  with  learning  in humans   (or  other animals)
entrained  in a systematic activity such   as a repetitive competition
game. In  the first place,  the mutual  search  for regularity  in the
opponent's behavior leads the whole system to develop highly irregular
evolution over long times,    which can hardly be distinguished   from
purely  random dynamics.   In   the  second   place, we   stress  that
competition can here be  interpreted as a  form of mutually supervised
learning and, thus,  results in a  kind  of collaboration between  the
opponents.

Some natural extensions of the present model are worth considering for
future  work. An important question to  be  addressed regards the case
where the entangled perceptrons are not equal  in size, i.e. they have
different numbers of neurons. In such a situation,  in fact, the above
quoted  correspondence of competition  and collaboration could fail to
hold.  Preliminary results  along  this line  (not  presented in  this
paper)  suggest however that the advantage  of  a larger perceptron is
relatively small.      Only very small  networks   ($N   \sim  2$) are
systematically defeated by larger opponents, as they typically fall in
short-period cyclic orbits.

The perceptron-like structure of our networks is probably the simplest
instance  among a large     class  of possible  architectures.   Fully
connected  networks  and multilayer structures     have been shown  to
exhibit     very     high     performance    in     learning     tasks
\cite{Hz,NN,Treves}. It  would therefore be   interesting to study how
these  more   complex networks      respond to  mutually    supervised
learning.  Finally, from the viewpoint  of  game  theory, it would  be
relevant to analyze the    dynamics of competing networks   engaged in
other games, especially,  when ordinary optimization procedures do not
lead to the optimal  playing strategy. We  mention, in particular, the
iterated prisoner's dilemma  \cite{Axel}, which is attracting  a great  
deal of  attention  as  a  paradigm  of      competition-collaboration
interplay,   and multiplayer minority games, recently studied by means 
of ensembles  of  globally  coupled  perceptrons  \cite{cm}. Competing  
neural networks could contribute  to  a  better  understanding  of the 
complex learning mechanisms involved in  such kind of social 
interactions.

\end{document}